\documentclass[10pt,twocolumn,preprintnumbers,amsmath,amssymb,nofootinbib
,superscriptaddress]{revtex4-1}
\usepackage{graphicx,longtable,mathrsfs,color,array}
\usepackage[hidelinks]{hyperref}
\usepackage[usenames,dvipsnames]{xcolor} 
\usepackage{amssymb,amsmath,mathtools,mathrsfs,slashed} 
\usepackage{epsfig,subfigure,placeins,float} 
\usepackage{booktabs,longtable,ctable,multirow} 
\usepackage{exscale,relsize} 
\usepackage[normalem]{ulem} 
\usepackage{enumerate}
\usepackage{times, mathptmx} 
\usepackage[utf8]{inputenc}
\usepackage{color}
\usepackage{hyperref}
\usepackage{graphicx}
\usepackage{color}
\usepackage{graphicx,graphics}
\usepackage{bm}
\usepackage{tabularx}
\allowdisplaybreaks[1]
\graphicspath{{figures/}}

\begin{document}

\title{On the growth of massive scalar hair around a Schwarzschild black hole}
\author{Katy Clough}
\email{katy.clough@physics.ox.ac.uk}
\affiliation{Astrophysics, University of Oxford, DWB, Keble Road, Oxford OX1 3RH, UK}
\author{Pedro G. Ferreira}
\email{pedro.ferreira@physics.ox.ac.uk}
\affiliation{Astrophysics, University of Oxford, DWB, Keble Road, Oxford OX1 3RH, UK}
\author{Macarena Lagos}
\email{mlagos@kicp.uchicago.edu}
\affiliation{Kavli Institute for Cosmological Physics, The University of Chicago, Chicago, IL 60637, USA}

\date{Received \today; published -- 00, 0000}

\begin{abstract}
Through numerical simulations of a minimally coupled massive Klein-Gordon scalar field, we show that it is possible to grow hair on a Schwarzschild black hole if one assumes an initial periodically time-varying but spatially homogeneous scalar background. By ``hair'', we mean a non-trivial profile in the scalar field. We find that this profile emerges on a timescale related to the mass of the black hole, with features related to the mass of the scalar particle. We undertake simulations with and without backreaction on the metric and see that the essential, qualitative features remain consistent. We also contrast the results from higher mass scalars to the case of a low mass with a large Compton wavelength. The results are particularly relevant for scalar-tensor theories of gravity and dark matter models consisting of a massive scalar, e.g.\ axions.
\end{abstract}

\maketitle


\section{Introduction}

Black holes are some of the most remarkable objects in nature. As endpoints of gravitational collapse they are where our current theory of gravity, General Relativity, is pushed to its extremes. 
While we have acknowledged their existence for many decades, the recent detection of black hole mergers by Advanced LIGO \cite{LIGO} and VIRGO \cite{Virgo}, or the imaging of the event horizon of the black hole at the heart of M87 \cite{2019ApJ...875L...1E}, further reinforce our belief in them as a part of the astronomical zoo.

Even though black holes are such exotic objects, they are also surprisingly featureless. In the context of General Relativity, and under certain assumptions, we know that a black hole is completely characterized by its mass $M$, spin $J$, and electromagnetic charge $Q$ \cite{Israel:1967za,Carter:1971zc}. This means that two black holes which have the same $M$, $J$ and $Q$ are completely indistinguishable. Furthermore, even for a broad class of extensions of General Relativity (in particular, where the extensions have a cosmological impact), black holes maintain this featureless characteristic  \cite{Hui:2012qt,2014PhRvD..89h4056G,Sotiriou:2015pka,Tattersall:2018map} and cannot support a non trivial configuration in surrounding fields. For this reason, it is often stated that black holes have {\it no hair}. If we restrict ourselves to an electrically neutral, non spinning, spherically symmetric black hole, the spacetime line element is then given by:
\begin{eqnarray}
ds^2=f(r)dt^2-f^{-1}(r)dr^2+r^2(d\theta^2+\sin^2\theta d\phi^2) \label{Sch}
\end{eqnarray}
where $t$, $r$, $\theta$ and $\phi$ are Schwarzschild polar coordinates, $f(r)=1-\frac{R_s}{r}$ with $R_s=2GM$ being the Schwarzschild radius given only in terms of the mass of the black hole $M$ and Newton's constant $G$.
\begin{figure}[ht]
	\centering
	\includegraphics[scale=0.48]{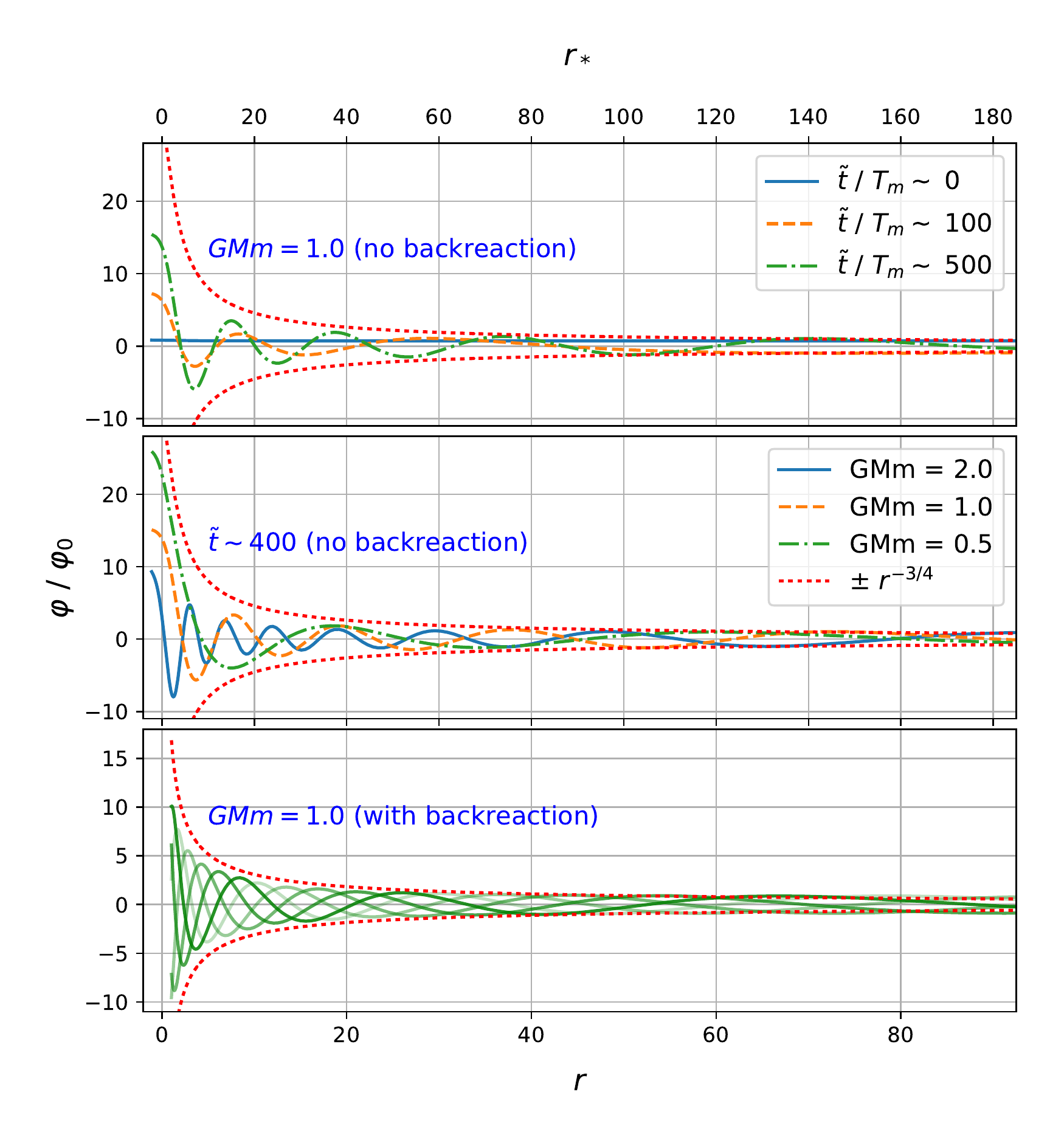}
	\caption{Two top figures show $\varphi$ as a function of $r_*$ without backreaction. The first plot illustrates the growth in time of the amplitude from a homogeneous state (with $T_m = 2\pi/m$). The second shows that the profiles depend on mass in a similar way to the expected analytical profiles -- i.e.~there are more oscillations for higher masses, and the envelope of the oscillations is well fit at larger radii by $r^{-3/4}$. The bottom plot shows a timelapse of $\varphi$ versus $r$ over a period of oscillation at a time of $\tilde{t}/T_m \sim 200$ for the backreacting case, which shows a consistent build up and profile.
	}
	\label{Fig:PhivsR_Timelapse}
\end{figure}

One might hope that, given their extreme nature, black holes could be used as a novel arena in which to explore new aspects of physics, but the expectation of having no hair works against this; if the black hole cannot support a non trivial environment, there will not be new physics to test. Fortunately, the expectation of featurelessness arises from a set of {\it no-hair theorems} which make a number of restrictive assumptions about the existence and environment of the black hole: that they are stationary, embedded in an asymptotically flat spacetime and that any new field, or ``hair" is regular at the horizon. Such conditions may be far from what a run-of-the-mill black hole will experience: the universe is expanding and full of material, much of which is not well understood.  We have learned from LIGO that many black holes do not exist in isolation, but may spiral into each other and merge in a highly dynamical and energetic process. In other words, black holes live in a universe which is not stationary or asymptotically flat. Thus, given our new found ability to probe black holes with gravitational waves or imaging, one needs to revisit the possibility that they may have hair in this context.

We know that many black holes do have long lived ``hair'' in the form of accretion discs of visible matter (although this is of course not stationary), but from a theoretical perspective, the possibility that a black hole may have scalar hair, $\varphi(r,t)$ is of particular interest. In this case, it will have a scalar field with a non trivial profile surrounding it which may, in turn, modify the spacetime metric (see \cite {Herdeiro:2015waa} for a review). Scalar fields exist in nature and arise as the low energy degree of freedom for a number of candidate fundamental theories  \cite{Clifton:2011jh, 1974IJTP...10..363H,Deffayet:2009wt} where they may be non-trivially (or non-minimally) coupled to the metric. In that case, they can be interpreted as new, ``fifth" forces which can be constrained in a number of ways \cite{Adelberger:2003zx}. Scalar fields are also the basis for several popular dark matter (DM) candidates, e.g.\ as an approximation to an axion-like particle \cite{1983PhRvD..28.1243T, Hu:2000ke, Amendola:2005ad, Marsh:2010wq}. 

There are reasonably general proofs that, under the restrictive conditions described above, it is not possible to have scalar hair for either minimal or non-minimal couplings. However, a number of counterexamples have been proposed for asymptotically flat spacetimes which lead to either long lasting hair or wigs  \cite{Cardoso:2011xi,Burt:2011pv,Cardoso:2016ryw, Herdeiro:2014goa,Ternov:1978gq}, by relaxing one or more of the no hair assumptions. For example in the case of Herdeiro and Radu's hairy black holes \cite{Herdeiro:2014goa} the complex scalar field around a spinning black hole varies in time (although the stress energy tensor is constant) and the solution is characterised by a Noether charge from the scalar sector, in addition to the usual BH parameters. Another important case also occurs for rotating -- i.e.\ Kerr -- black holes, where a superradiant instability can lead to a copious production of scalar particles by amplification of quantum fluctuations in a bosonic field (see \cite{Brito:2015oca} for a review). The particles extract angular momentum from the BH to populate bound states, leading to clouds which decay over much longer periods into gravitational radiation. In this paper we will discuss another possibility: that the growth of hair may result from the black hole being embedded in a cosmological spacetime, containing an asymptotically time-varying but spatially homogeneous scalar field.

In a seminal paper \cite{Jacobson:1999vr}, Jacobson showed that it was possible to endow a black hole with hair if it is embedded in a cosmological spacetime dominated by a massless scalar field. Specifically, he showed it was possible to construct a non-trivial solution for $\varphi(t,r)$ which varied with time asymptotically, $\varphi(t,r=\infty)\propto t$ and which was regular at the horizon. Jacobson's solution is a proof of concept that it is possible to endow a black hole with hair by mildly violating the conditions for the no-hair theorems. The amplitude of the hair is proportional to ${\dot \varphi}$ at $r=\infty$; if ${\dot \varphi}\sim H_0$ (Hubble rate today) and monotonic, as one might expect in a cosmological setting, then the effect of hair will be negligible. The possibility remains, however, that ${\dot \varphi}\gg H_0$ which would lead to a more substantial effect, and it is this scenario which we investigate here.

We will focus on the case of a single massive scalar field minimally coupled to the metric. This field will satisfy the Klein-Gordon equation on a curved spacetime. For concreteness, let us consider a scalar with a potential $V(\varphi)=\frac{1}{2}m^2\varphi^2$, which could be the dominant term in the Taylor expansion of a more general potential arising, for example, in general scalar-tensor theories considered in the Einstein frame \cite{1974IJTP...10..363H,Deffayet:2009wt} or could be a good approximation to a light bosonic particle with mass $m$. 

There is extensive work on the role massive scalar fields can play in cosmology \cite{Turner:1983he,Marsh:2010wq}. If the field oscillates coherently, it behaves (on average) like a homogeneous source of dark matter; inhomogeneous modes will cluster, with an emergent sound speed set by the Compton wavelength of the scalar field, $\lambda_C\sim 1/m$. Particularly in low mass cases, scalar particles may form halos which are partially stabilized by their macroscopic quantum behaviour, leading to distinct signatures. 

Here, we numerically show, for the first time, how an initially homogeneous and isotropic cosmological scalar field, with $\varphi(t)\propto e^{imt}$, evolves in time and settles around a spherically-symmetric BH, forming a scalar cloud (illustrated in Fig.~\ref{Fig:PhivsR_Timelapse} with detailed discussion in Section \ref{Results}). Furthermore, we do so self consistently, including the backreaction of the scalar field on the metric.

In the limit of large $m$, what we are simulating describes the infall and clustering of the scalar particles around the black hole. We find that a non trivial profile grows on a timescale related to the mass of the black hole (its freefall timescale), with features in the cloud related to the mass of the scalar particle. Note that this mechanism is very different from the formation of superradiant clouds -- the energy of the cloud is fed by the asymptotic oscillations of the field, not from the black hole itself (which in this work is non spinning). Since the field at the boundary is non zero, and homogeneously oscillating, it provides an infinite reservoir of particles with which to feed the black hole. 

We structure the paper as follows. In Section \ref{Framework} we describe the details of the system we are studying and the numerical implementation which is employed. In Section \ref{Results} we present the results for a number of situations: without and with the backreaction on the metric evolution and for infalling and standing waves in the scalar field. In Section \ref{Discussion} we discuss our results and the physical interpretation of the new effect that we have studied.

Note that our simulations use geometric units in which $G=c=1$, and thus the scalar field is parameterised by the inverse length scale $\mu$, expressed in terms of the Schwarzschild radius $R_s$.  In this paper we relate this length scale to the scalar mass via $\mu = m c / \hbar$ and restore the $G$ in some places so that the units are consistent. Thus our simulations are parametrised in terms of the dimensionless quantity $GMm$, which then sets the mass of the black hole if the mass of the particle is specified, and vice versa. As an example, for the case $GMm=1$, this could correspond to a solar mass BH and a scalar with mass $m \sim 10^{-10}$ eV, or a supermassive BH of $M= 10^{10}M_\odot$ with a scalar mass $m \sim 10^{-20}$ eV.



\section{Framework and Numerical Setup}
\label{Framework}
To begin with, let us consider the case where Eq.\ (\ref{Sch}) is a background metric on which the dynamical scalar field evolves (i.e.\ no backreaction). In this case, the Klein-Gordon equation on a Schwarzschild background gives the following evolution equation for a spherically-symmetric scalar $\varphi(t,r)$ \cite{Regge:1957td, chandrasekhar1998mathematical}:
\begin{eqnarray}\label{KGeqn}
\left[\frac{d^2}{dt^2}-\frac{d^2}{dr^2_*}+V_Z\right]r\varphi(t,r)=0,
\end{eqnarray}
where we have introduced the tortoise coordinate $r_*= r + R_s\ln(r/R_s-1)$ and the effective potential:
\begin{eqnarray}
V_Z(r)=f(r)\left(\frac{R_s}{r^3} + m^2\right).
\end{eqnarray}
One can look for stationary solutions of the form $\varphi=e^{i\omega t}\psi(r)$. Ingoing boundary conditions are imposed at the horizon, due to the causal structure of the black hole. However, we allow both ingoing and outgoing modes at spatial infinity (which allows the temporal frequency $\omega$ to take real values). This choice of boundary conditions mimics the effect of sources external to the black hole which, in our case, describe the cosmological background in which the black hole is embedded. This background acts as a reservoir, from which particles fall into the black hole and create an ingoing flux of waves that cause the formation of a cloud surrounding the black hole.
With this choice of boundary conditions, it is possible to construct exact solutions for the entire spacetime \cite{Fiziev:2005ki,Fiziev:2006tx,Bezerra:2013iha,Vieira:2014waa}, based on the Heun functions which describe a scalar cloud surrounding the BH\footnote{Note that these solutions are fundamentally different to the bound states which are excited by superradiant growth.}. For a thorough analysis of the scalar field cloud in this case, see \cite{Lam2019}. There will be two qualitatively different situations to distinguish: when (i) $R_s m< 1$ and (ii) $R_sm\gtrsim  1$. Generically, the solution to the equations will allow for frequencies $\omega^2= m^2+k^2$, with $k$ denoting the spatial energy of the scalar field at infinity in the Schwarzschild geometry. 
As we will discuss later, due to the homogeneous initial profile chosen for the scalar, the numerical solutions described in this paper will fit profiles with $\omega=m$. For this reason, from now on we limit ourselves to this choice.
On the one hand, in case (i), the Compton wavelength of the scalar particle is outside the horizon and it creates a potential barrier that will reflect ingoing waves near the horizon. In this case, whereas the near-horizon solutions ($r_*\rightarrow -\infty$ or $r\rightarrow R_s$) are given by $\varphi \propto \exp\{ -im(t+r_*-r)\}$ for $\omega=m$, far from the horizon we have standing waves given by $\varphi \propto \exp\{  -imt\}\cos(2m\sqrt{R_sr})/r^{3/4}$. On the other hand, in case (ii), the Compton wavelength of the scalar particle is inside the horizon and the potential becomes purely attractive, leading thus to ingoing waves only. The solutions near the horizon are given by  $\varphi  \propto e^{i \omega(t+r_*)}$ whereas far from the horizon by $\varphi \propto  \exp\{-i m\left( t+ 2\sqrt{R_s r} \right) \}/r^{3/4}$ for $\omega=m$. 


We emphasize that these analytical solutions characterize stationary scalar fields with an amplitude reaching a maximum around the black hole and decaying at infinity. However, in this paper we focus instead on the dynamical growth of hair from a completely homogeneous scalar field to the cloud previously mentioned. In addition, we consider a somewhat different setup, one of a cosmological background in which the field has an asymptotically finite amplitude of oscillation, at least over some region large relative to $R_s$. Nevertheless, as we will discuss in the following sections, we find that the scalar profile close to the BH shares the features of the stationary analytical solutions at any moment during the growth. 

\subsection{Metric setup: without backreaction}

For cases without backreaction, we employ the code and methods of \cite{Alexandre:2018crg}. We fix the metric to Eq.~(\ref{Sch}), although for practical reasons we use Cartesian Kerr-Schild coordinates (see e.g.\ \cite{Witek:2012tr}), which are horizon penetrating. The Kerr-Schild time coordinate ${\tilde t}$ is related to the Schwarzschild coordinate $t$ through $\tilde{t} = t + R_S \ln \left(r/R_S - 1\right)$, and the radial coordinate is that of Schwarzschild (in presenting our results we convert to the tortoise coordinate $r_*$ for clarity). In such coordinates the metric in the standard  $3+1$ ADM decomposition is given by:
\begin{equation}
ds^2=-\alpha^2\,dt^2+\gamma_{ij}(dx^i + \beta^i\,dt)(dx^j + \beta^j\,dt),
\end{equation}
with components
\begin{equation}
\alpha = \frac{1}{\sqrt{(1+2M/r)}},\; \beta^i = \frac{2 M x^i}{r + 2 M},\; \gamma_{ij} = \delta_{ij} + \frac{2 M x_i x_j}{r^3} ,
\end{equation}
where $x^i=x_i$ are the Cartesian coordinates on the grid and $r^2 = x^2 + y^2 + z^2$.
The trace of the extrinsic curvature $K$, the only other component required for the evolution of the fields, is given by:
\begin{equation}
K = 2 \alpha^3 (1+H) x^i \partial_i H + 2 \alpha H \partial_i (x^i/r),
\end{equation}
where $H = M/r$.

The field evolves on this background according to the two first order evolution equations
\begin{align}
\partial_t \varphi &= \alpha \Pi +\beta^i\partial_i \varphi \label{eqn:dtphi} ~ , \\ 
\partial_t \Pi &= \alpha\partial_i\partial^i \varphi +\alpha\left(K\Pi -\gamma^{ij}\Gamma^k_{ij}\partial_k \varphi +\frac{dV}{d\varphi}\right)\nonumber  \\
& + \partial_i \varphi \partial^i \alpha + \beta^i\partial_i \Pi \label{eqn:dtPi} ~ .
\end{align}

The metric necessitates excision at the singularity -- in the static Schwarzschild case this can be done by simply setting the field within the horizon (in practice we do this for $r < R_s/ 2$) to decay to zero. Since the curvature of the metric prevents signals from escaping, errors due to the excision do not in principle propagate outwards. 

In this fixed case, we set $R_s = 1$ and study the formation of the cloud in four different cases: $GMm=0.1$, $GMm=0.5$, $GMm=1$, $GMm=2$. The first case represents a ``low mass'' case where the effect of the pressure from the Compton wavelength of the scalar is significant. In the remaining cases, this effect is smaller with the field behaving more like a pressureless dust as $GMm$ increases. We refer to these then as the ``high mass'' cases.

\subsection{Metric setup: including backreaction}

For cases with backreaction, the full numerical relativity capabilities of the publicly available numerical relativity (NR) code $\textsc{GRChombo}$ \cite{Clough:2015sqa, Chombo}  (\url{www.grchombo.org}) are used. The specific approach and code was adapted from that used in \cite{Clough:2018exo}, which employed standard Numerical Relativity (NR) techniques. The BSSN/CCZ4 formulation of the Einstein equations is combined with the moving puncture method for stable evolution of BH spacetimes, and the initial data is conformally flat Bowen-York data \cite{BowenYork} for an isolated (non spinning, non boosted) BH. We are required to add a spatially constant value for the trace of the extrinsic curvature, $K = - \sqrt{24\pi \rho_0}$, so that the Hamiltonian constraint
\begin{equation}
\mathcal{H} = R + K^2-K_{ij}K^{ij}-16\pi \rho = 0 
\end{equation}
is properly satisfied in the presence of the non-zero field. Here, $\rho$ is the energy density of the scalar field with initial value $\rho_0$. This results in a background with the spacetime asymptotically expanding, as motivated by our cosmological scenario. For these dynamical metric simulations we focus only on the case with $GMm=1$, since we expect the effects to be consistent in the other cases.

\subsection{Initial conditions, numerical methods, and boundary conditions}

For all simulations the initial conditions for the scalar field are set to $\varphi( t=0,r)=\varphi_0$ and $\partial_t \varphi (t=0,r)= 0$. 
We choose an initial amplitude $\varphi_0$ such that the energy density is $\rho( t=0,r)=\rho_0 = 10^{-10} R_s^{-2}$ in geometric code units. In physical units this is {\it significantly} higher than the average energy density of dark matter\footnote{$10^{-10} R_s^{-2}$ is between $\sim 10^{30}$ greater than the DM density in the case of solar mass BHs (equivalent to the density of a white dwarf star) and $\sim 10^{10}$ times greater for SMBHs like that of M87 (equivalent to a density of $10^{-10} ~ \rm{kg/m^3}$). Note that similarly unrealistic values are sometimes used in evolving Neutron Star binaries in a non zero atmosphere.}. In the case without backreaction, we are only interested in the evolution of the field and the absolute energy density of the field is not particularly relevant, but in the case with backreaction we should bear in mind that the expansion of spacetime and backreaction effects will be significantly higher than in a physically realistic scenario, so this is an overly stringent test of robustness to backreaction effects.

The numerical methods are those of $\textsc{GRChombo}$, being the method of lines with finite difference stencils, Runge Kutta time integration and a hierarchy of grid resolutions. Note that in the fixed background case the metric does not evolve dynamically, and the metric components are calculated analytically rather than stored on the grid. The size of the domain is $L=1024R_S$, and we use seven (2:1) refinement levels with the coarsest having $256^3$ grid points, although we use the octant symmetry of the problem in Cartesian coordinates to reduce the domain to $128^3$ points. The fact that we do not explicitly impose spherical symmetry means that in the future we will be able to easily generalize these simulations to spinning and binary black holes. However, it also breaks the spherical symmetry of the domain, which may explain the late time instabilities in the evolution which are discussed below.

We implemented non-zero, time oscillating boundary conditions for the scalar field by extrapolating the field from within the numerical domain. After testing several possibilities, we imposed that the field amplitude is extrapolated either linearly $\varphi = B - Ar$ or as $r^{-3/4}$ at the boundaries, with the case chosen at each step which minimizes the change from the values within the domain. During the build up of the cloud the energy density at the boundaries decreases only fractionally and the field profile remains smooth as it oscillates. Note that, unlike in the analytical solution, our boundaries do not permit ingoing waves of frequency $m$, and are not initially constrained to decay as $r^{-3/4}$. Despite these differences, we see the growth of a cloud with a similar form to the analytical solutions.

\begin{figure}[h]
	\centering
	\includegraphics[scale=0.4]{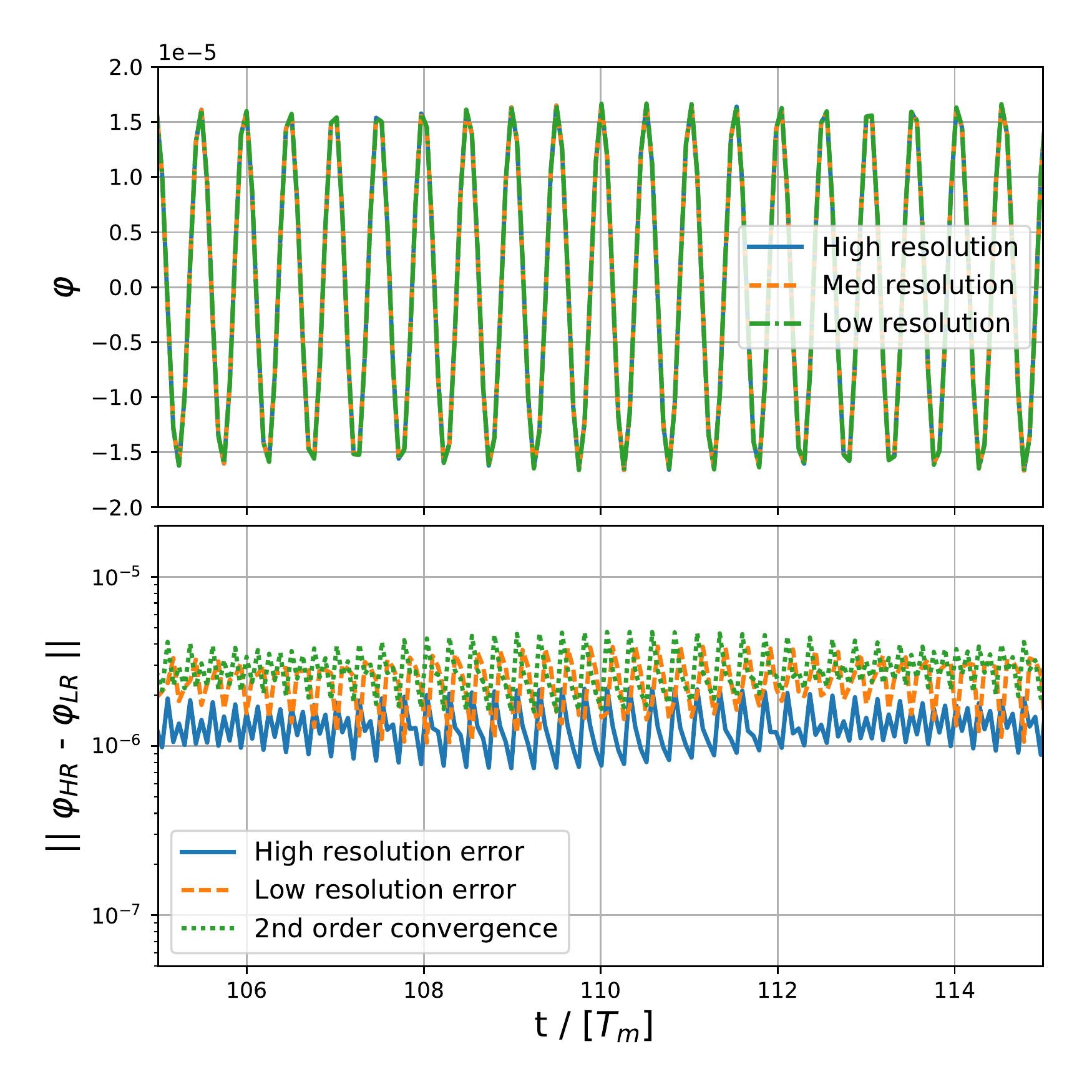}
	\caption{Plot showing an example of the tests performed to validate the convergence of the code. We check that the evolution of the field is indistinguishable at different resolutions, and that the errors converge as resolution is increased (here at second order).}
	\label{Fig:ConvergenceTest}
\end{figure}

For all our setups we perform validation tests of the metric used, check the Hamiltonian constraint violation is bounded and decreases with resolution (for backreacting cases), and test convergence of the field as illustrated in Fig. \ref{Fig:ConvergenceTest}. The numerical errors over the periods for which the results are presented are not significant.

\section{Results}
\label{Results}

\subsection{Fixed background metric without backreaction}
We begin with the results without backreaction, in which the scalar field is evolved on a fixed Schwarzschild background. 
In Fig.\ \ref{Fig:PhivsR_Timelapse} we see a steady growth of the scalar amplitude around the BH for $GMm=1$, starting from a homogeneous profile; after an initial transient period, there is an approximately linear build up in the ingoing waves, which have a spatial envelope well fit by a $1/r^{3/4}$ dependence, and a spatial wavelength which decreases with increasing $m$. Note that further out the field asymptotes to a homogeneous profile (the boundary is at $r=512R_s$), so an extrapolation of the $1/r^{3/4}$ fit at smaller $r$ lies above the asymptotic values. Over time the amplitude at the boundary is decreasing, but only very slightly -- as explained further below, it appears that the field is gradually redistributing itself into the analytic profile which matches the amplitude imposed at the boundary. 

\begin{figure}[t]
	\centering
	\includegraphics[scale=0.35]{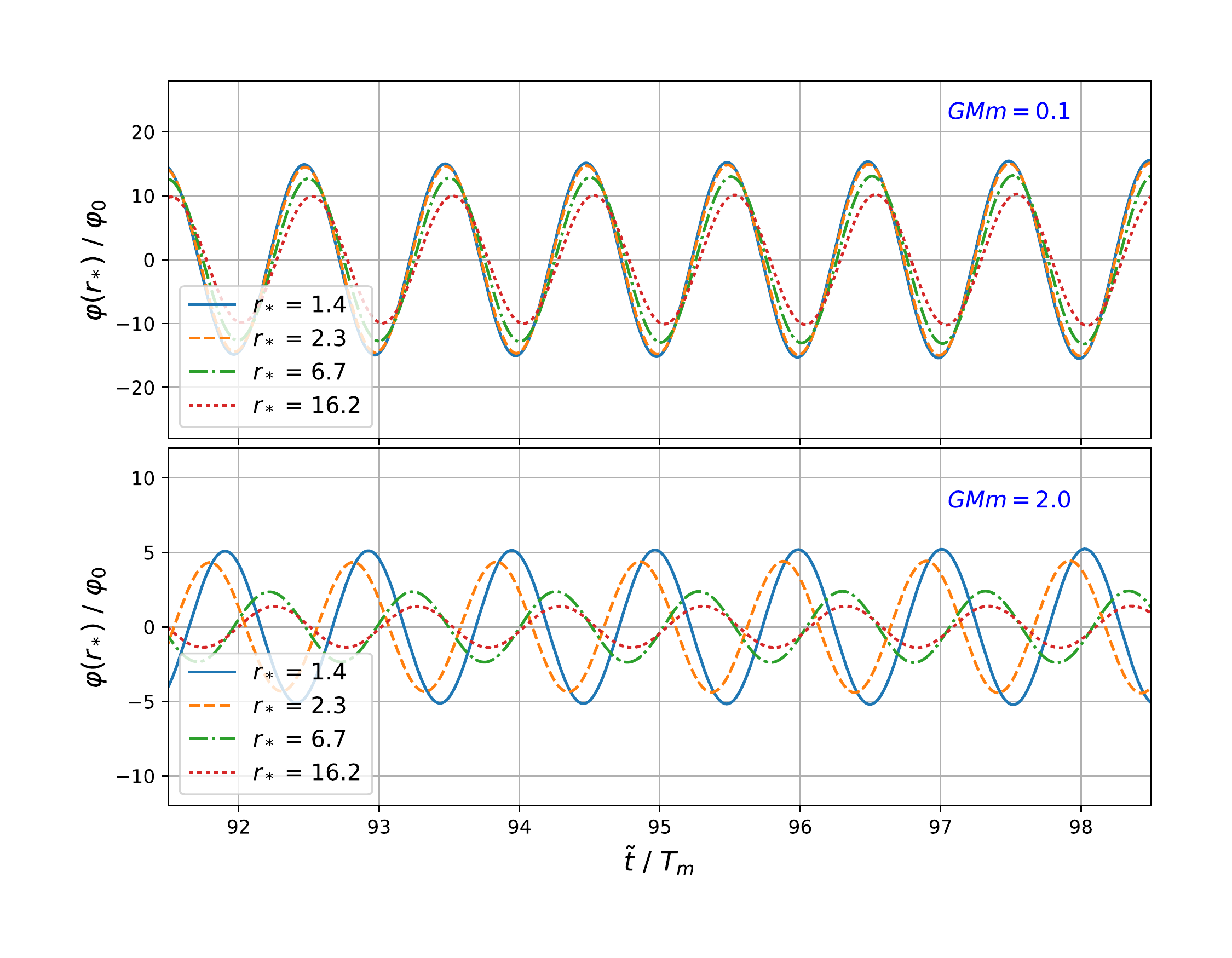}
	\caption{Plot showing frequencies at different radii in the stationary wave $GMm=0.1$ and ingoing wave $GMm=1.0$ cases. In the standing wave case the frequencies are synchronised, whereas in the infalling case the frequencies are independent of radius, but are out of phase.}
	\label{Fig:Frequencies}
\end{figure}

\begin{figure}[t]
	\centering
	\includegraphics[scale=0.5]{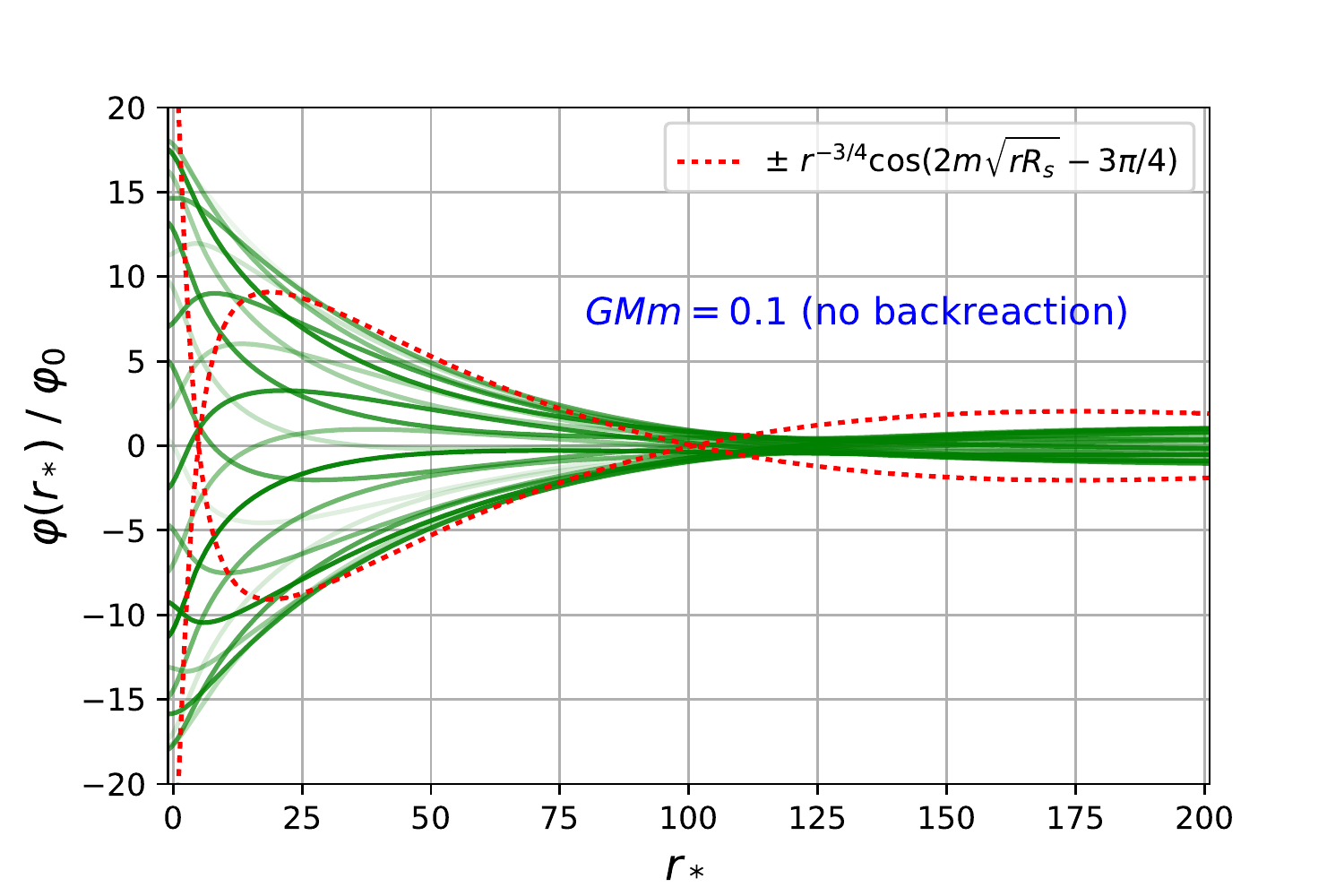}
	\caption{A timelapse of $\varphi$ versus $r$ over a period of oscillation at a time of $\tilde{t}/T_m \sim 100$ for the non-backreacting case with $GMm=0.1$. In this small mass regime a standing wave profile is observed rather than an inflow. The dotted envelope is the profile expected from the analytical solutions in the case of $k=0$.
	}
	\label{Fig:PhiStandingWave}
\end{figure}

During the initial growth of the cloud, the temporal frequency of the oscillations depends on time and space, but after only $\sim 20$ oscillations it converges to a constant given by $m$. We observe a similar behaviour for the spatial frequencies, with a spatial dependence on $m$ as shown in Fig.~\ref{Fig:Frequencies}. Our results thus approximately fit the $k=0$ analytical results, which is consistent with our choice of initial conditions -- set such that the spatial momentum of the field is zero. Also, this is consistent with the cold DM case, for which the momentum is small relative to the mass (although one would expect a non zero angular momentum). In the low mass regime, $GMm \ll  1$, as explained above, we expect standing waves rather than ingoing waves, and this is consistent with what we observe from the frequencies for the case $GMm=0.1$ in Fig.~\ref{Fig:Frequencies}, and also from the profiles shown in Fig.\ \ref{Fig:PhiStandingWave}.


\begin{figure}[b]
	\centering
	\includegraphics[scale=0.5]{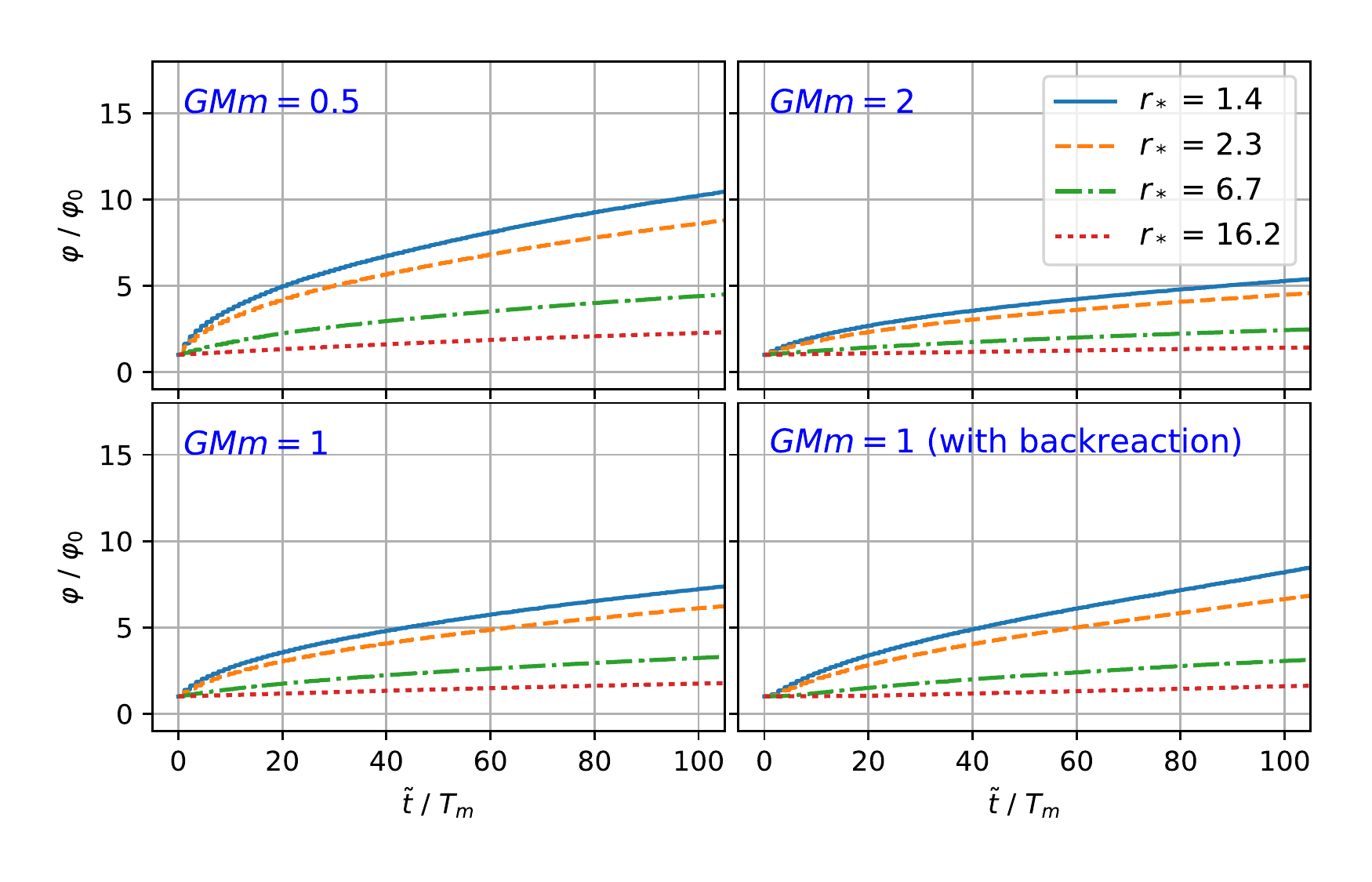}
	\caption{The amplitude of the oscillations in $\varphi$ as a function of the number of oscillations $T_m = 2\pi/m$ for various radii and values of $GMm$ as labelled. The bottom right figure shows the case of $GMm=1.0$ with backreaction included. Note that in the backreaction case we cannot define $r_*$ as the gauge is dynamical, but we select the same coordinate locations as the fixed case for the plot. Whilst these will not give the same physical positions in space, we still see a consistent pattern for the build up.}
	\label{Fig:PhivsTime}
\end{figure}

We show the time evolution in more detail in Fig.\ \ref{Fig:PhivsTime} for the high mass cases, where the amplitude of the scalar field oscillations at several values of $r_*$ are plotted versus the number of oscillations in the field. In these cases where $GMm \sim 1$, the rate of growth of the field amplitude as a function of the number of oscillations, $n$,  is faster than in the lower mass cases, with a dependence on $m$ after the initial transient period of $\frac{1}{\varphi_0} d\varphi/dn \propto 1/\sqrt{m}$. Given that the oscillation period $T_m \propto 1/m$, the physical rate of build up of the field profile will scale as $\frac{1}{\varphi_0} d\varphi/dt \propto \sqrt{m}$.

\begin{figure}[b]
	\centering
	\includegraphics[scale=0.40]{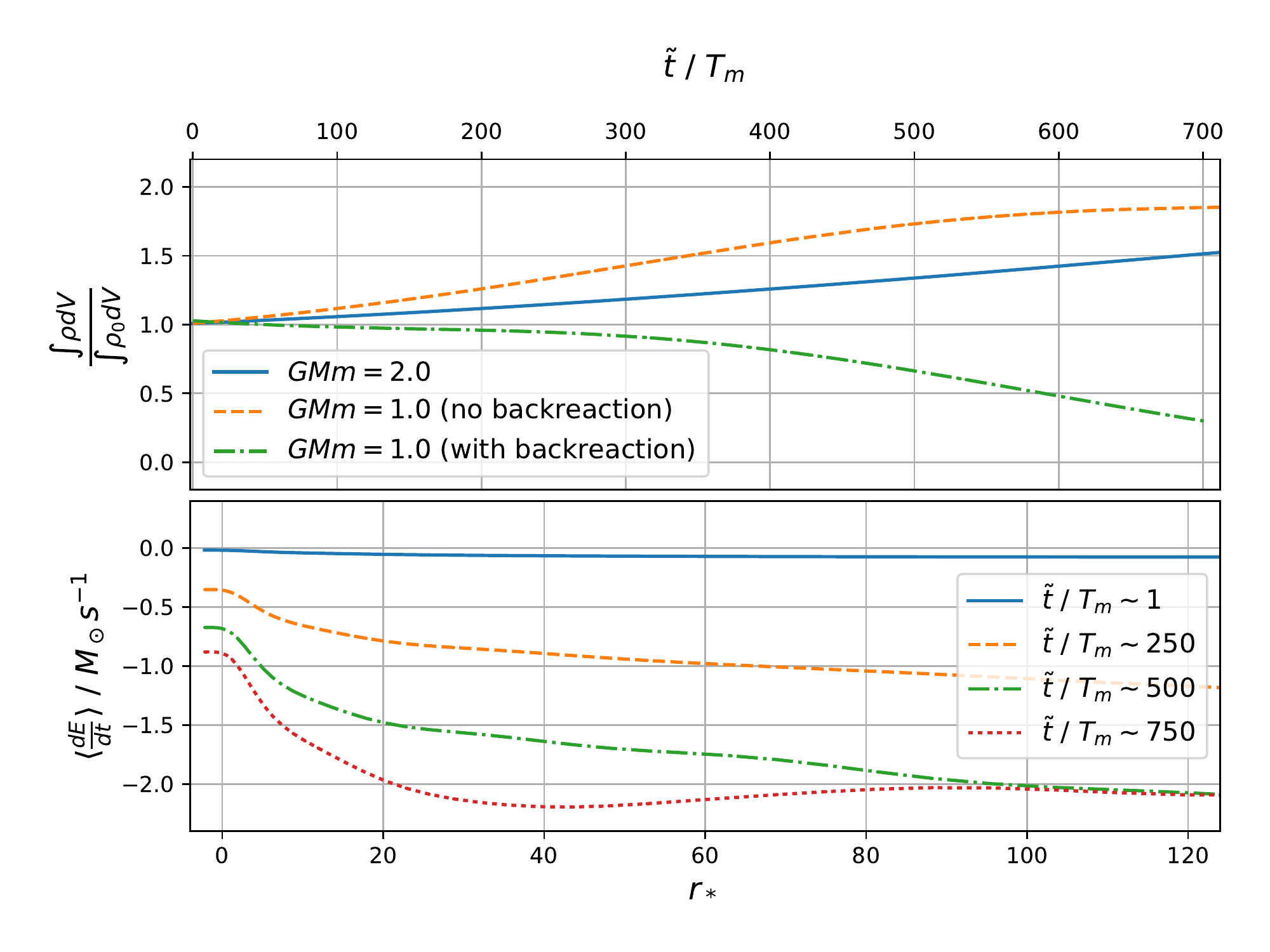}
	\caption{The top panel shows the volume integral of $\rho$ within a radius $r=100R_s$ (excluding the region within the horizon) as a function of $T_m = 2\pi/m$. In the backreaction case an energy decrease is observed due to the effect of the cosmological expansion, but the other cases consistently show a growth. There are small oscillations in the growth on the timescale of the mass but these are not visible in the plot. The bottom panel shows the energy flux $dE/dt$ across spheres of radius $r_*$, averaged over several periods, for the case $GMm=1$ without backreaction. Note that the relatively high flux values are the result of the large value of $\rho_0$ used in our simulations, and one would expect the actual rates to scale with $\rho_0$ accordingly.}
	\label{Fig:RhovsTime}
\end{figure}

The top panel of Fig.~\ref{Fig:RhovsTime} shows the volume integral of the energy density within a radius of $r=100R_s$ (excluding the region within the horizon), with $\frac{1}{\int \rho_0 dV} d(\int \rho dV)/dn \propto 1/m$ as expected since the energy density scales as $\rho \propto \varphi^2$. Thus the energy density grows at a rate in real time which is independent of $m$, and we measure the physical time for the mass within $r < 100R_s$ to double as $t = ( {M}/{M_\odot}) 6 ~ \rm{ms} $.
These values are consistent with the freefall timescale for the BH, which depends on $M$ but not $m$ whilst the cloud mass is negligible. This is reasonable as the oscillating scalar should behave roughly as matter for $GMm>1$, which simply falls in radially under the gravity of the BH (note that our initial conditions do not provide it with angular momentum for support). 
The bottom panel of Fig.~\ref{Fig:RhovsTime} shows the energy flux across spheres of radius $r_*$, calculated as $dE/dt = 4\pi r^2 (1+M/r)^6 S_r$ with the momentum density $S_r(r) \propto \partial_r\varphi\partial_t\varphi$. The values are averaged over several oscillations, showing a net inflow towards the BH (note that the momentum density actually oscillates between inflow and outflow over each period of scalar field oscillations). Initially the rate of inflow is roughly that of freefall in the regions exterior to the cloud, but suppressed at the horizon, which explains further why the cloud grows at approximately the freefall rate for the BH -- at least initially, the inflow into the BH can be neglected.
As the cloud approaches the analytic profile, the rate of infall at the BH horizon increases, tending towards that at larger radii, which is consistent with the idea that the solution will settle into a steady state with a rate of inflow which is independent of $r$. 

\begin{figure}[b]
	\centering
	\includegraphics[scale=0.40]{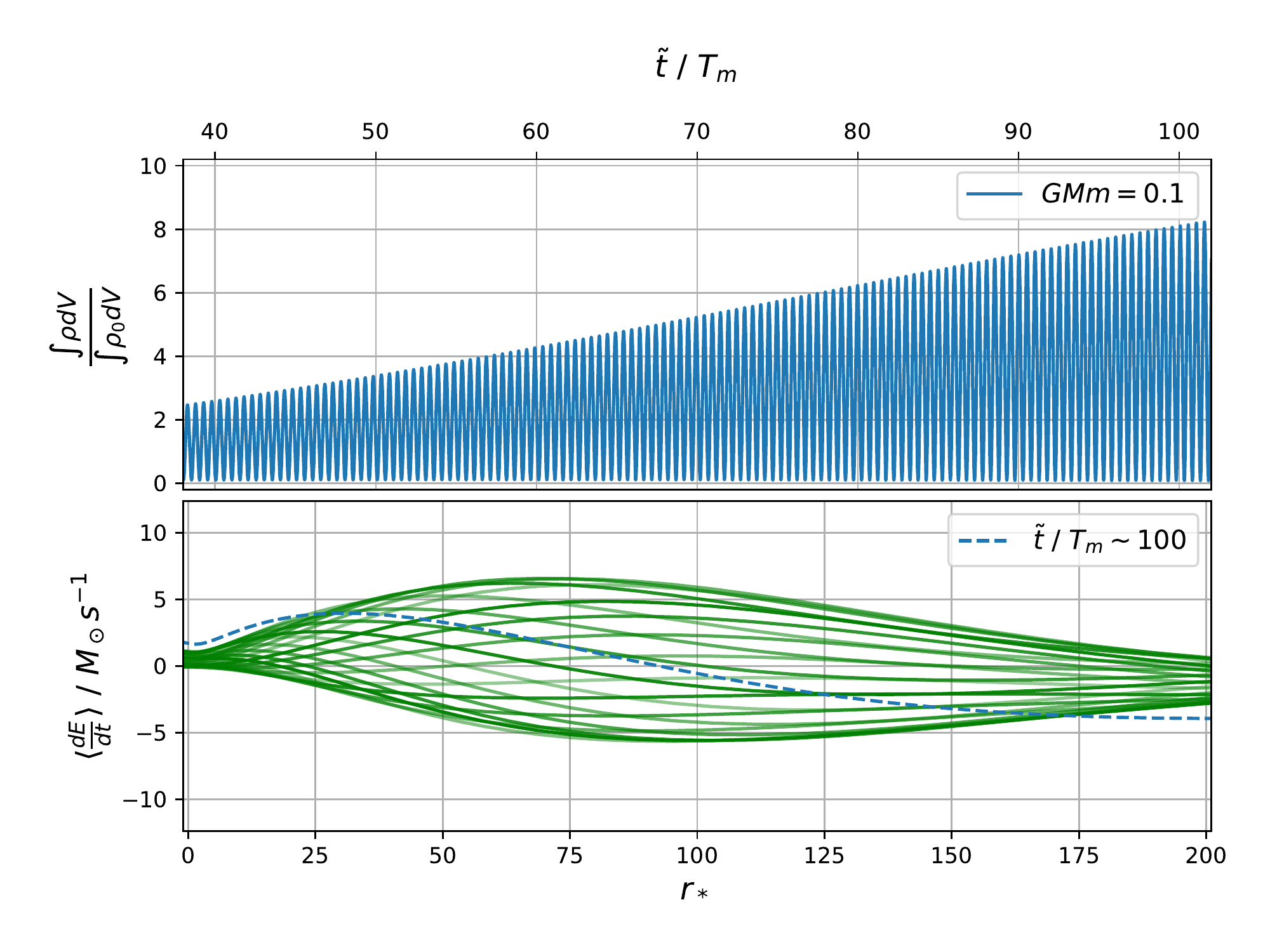}
	\caption{Both plots show the low mass case $GMm=0.1$ without backreaction. The top panel shows the volume integral of $\rho$ within a radius $r=100R_s$ (excluding the region within the horizon) as a function of $T_m = 2\pi/m$. In contrast to the higher mass cases shown in Fig. \ref{Fig:RhovsTime}, the value oscillates every cycle between a maximum and minimum value due to the standing wave nature of the solution. The rate of growth is also more rapid and faster than the freefall timescale. The bottom panel shows a timelapse of the energy flux $dE/dt$ across spheres of radius $r_*$, with the dotted line showing the value averaged over several periods. The point of zero average flow corresponds to the node see at $r_* \sim 100R_s$ in Fig. \ref{Fig:PhiStandingWave}. }
	\label{Fig:RhovsTimeSmallMass}
\end{figure}

Note that the growth behaviour is very different for cases where $GMm < 1$, where the analytic solutions change to standing waves and the timescales depend on $m$ due to the effect of the so-called ``quantum pressure'' from the comparably large Compton wavelength. This can be seen in the case $GMm=0.1$ that we studied, where there is a much more rapid build up in the cloud, with negligible inflow at the horizon (due to the reflection of the modes off the potential barrier), as shown in Fig.~\ref{Fig:RhovsTimeSmallMass}.


Regarding the final state of the build up - in the case of $GMm=1$ it appeared that the cloud began to saturate at a constant energy density around $\tilde{t} \sim 700 T_m$. However, in all cases we found that at later times the field oscillations became disturbed, with large oscillations in the amplitudes and no clear steady state. This is most likely due to unphysical effects from the imposed boundaries, which do not permit ingoing waves, and also (due to the cartesian grid) do not reflect the spherical nature of the solutions. This requires further investigation, but notice in Fig.\ \ref{Fig:PhivsR_Timelapse} that the spatial oscillations and $r^{-3/4}$ profile spread outwards from the BH as time elapses. At some point the boundary conditions start to be inconsistent with the physical solution, in which the oscillations would continue to spread outwards throughout the homogeneous field, resulting in an inflow at $r=L$. In the absence of these boundary effects, we would expect the cloud to settle into a steady state, with a constant inflow, and a density profile set by the value imposed at $r=L$.

\begin{figure}[h]
	\centering
	\includegraphics[scale=0.5]{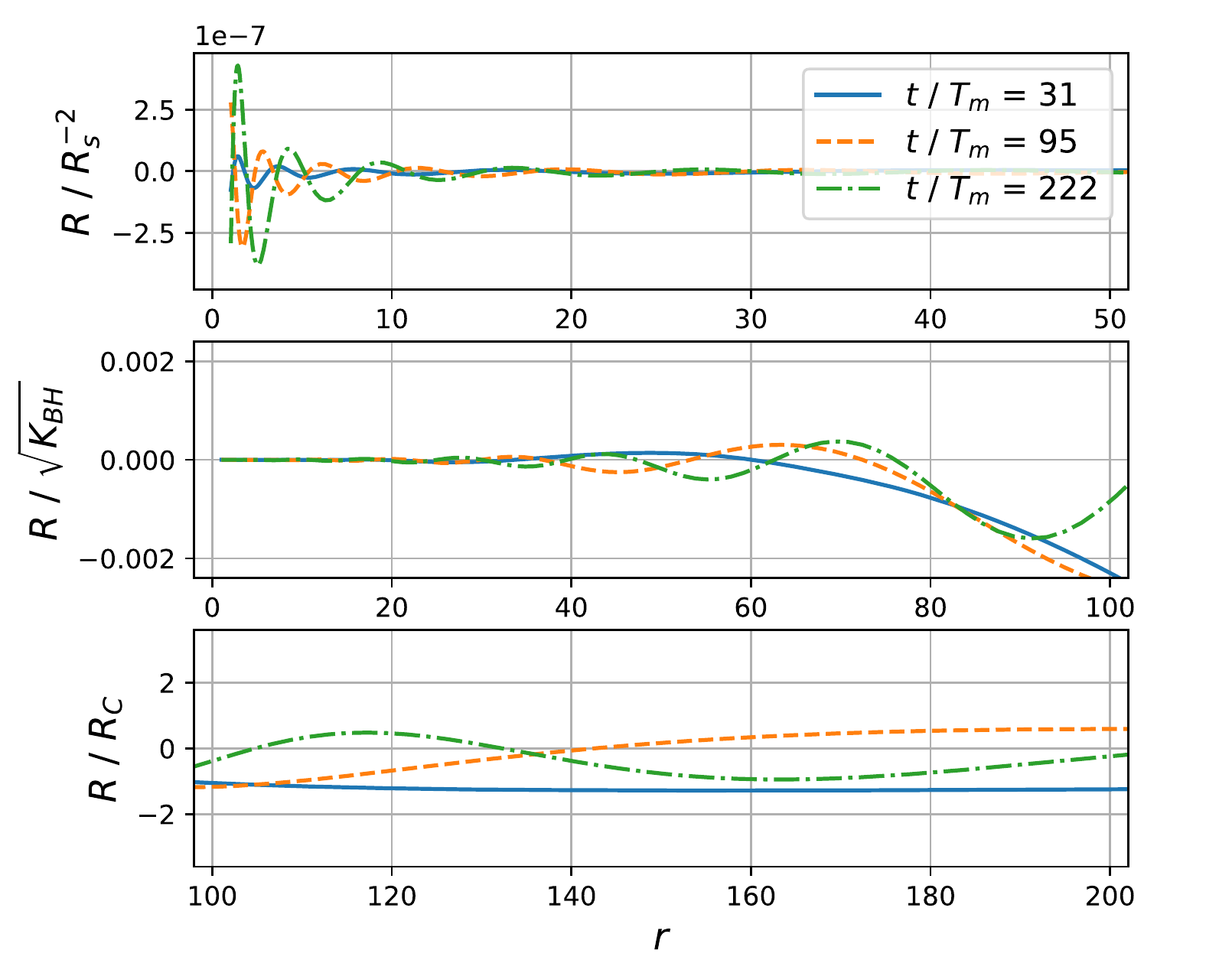}
	\caption{Plot showing the 4D Ricci scalar $^{(4)}R$ in absolute terms in units of $R_s^{-2}$ (top), relative to the scale of curvature of the BH $\sqrt{K_{BH}}$ at small radii (middle) and relative to the Ricci scalar for the asymptotic cosmological spacetime $R_C$ at larger radii (bottom). Whilst the effects close to the BH are small relative to the curvature of the BH itself, further out the deviations become more significant.}
	\label{Fig:RicciScalar}
\end{figure}

\subsection{Dynamical background metric with backreaction}
\noindent
We now focus on the case with backreaction, in which we simultaneously evolved the fully consistent metric and scalar field profiles. There are three key effects which are neglected in the non backreacting case, which we have taken into account here. Firstly, the accretion of the cloud onto the BH leads to a time varying mass which means that the analytic profiles discussed in \cite{Lam2019} evolve over time, albeit on a timescale much longer than the accretion rate. Secondly, due to the backreaction of the cloud on the metric, the background spacetime around the black hole is a perturbed Schwarzschild solution, showing imprints of the scalar density. Finally, the non zero asymptotic density results in an asymptotically de Sitter spacetime, with expansion that dilutes the cloud, rather than the asymptotically flat case. All of these effects could potentially disturb the build up of the scalar clouds.

The plots in Fig.~\ref{Fig:PhivsR_Timelapse} and Fig.~\ref{Fig:PhivsTime} compare the profiles and evolution of the scalar field as a function of time in the backreacting and non-backreacting cases. Whilst for backreaction the dynamical gauge means that the field cannot be directly compared at the same radial coordinate, the two do show a very similar rate of build up, meaning that, at least in the initial stages, backreaction does not suppress the growth of the cloud.

Whereas the build up shows roughly the same growth in the amplitude at a given coordinate radius in the backreacting case at a fixed tortoise coordinate $r_*$ as in the non-backreacting one, in the former the energy density of the cloud is actually being diluted by the expansion of the spacetime as the average scale factor increases by about 10\% during the simulation. Therefore, as illustrated in the top panel of Fig.~\ref{Fig:RhovsTime}, the overall energy of the cloud within a (coordinate) radius $r=100R_s$ decreases over time. This illustrates that, depending on the asymptotic energy density of the field, the dilution due to expansion plus the infall into the BH may out-compete the growth of the cloud. Partly this is due to the fact that the energy density of the cloud will scale as radiation rather than matter (due to the contribution of gradients). However, the value we have used of $\rho_0 \sim 10^{-10} R_s^{-2}$ in our simulations is orders of magnitude bigger than physically relevant scales, such as those in DM halos, and then only shows a small decrease. We thus conclude that the backreaction of the scalar on the metric and the effects of cosmological expansion should not disturb the build up significantly in realistic scenarios.

In Fig.~\ref{Fig:RicciScalar} we quantify the non-zero value of the 4D Ricci scalar as a result of the presence of the cloud, calculated as $^{(4)}R = 8\pi (\rho-S)$ with $S$ the isotropic stress of the scalar field. The absolute values are compared to the square root of the Kretchmann Scalar for the BH $K_{BH}=12R_s^2/r^6$ and the Ricci scalar for the cosmological spacetime $R_{C} = 24\pi G \rho_0$. Whilst the effects close to the BH are small relative to the curvature of the BH itself, further out the deviations become significant, and asymptote to values of order $R_C$, but with a characteristic oscillating imprint with a spatial scale related to $m$.

Whilst the effects studied in the backreacting case here are small in physically realistic scenarios, the effects may be enhanced by the addition of angular momentum, or in binary cases where resonances can occur in the scalar mass and the orbital period. These simulations therefore represent a significant step in going beyond the static background approximation, which is necessary if one wants to study the impact on the gravitational wave signal from DM environments in such cases.

\section{Discussion}
\label{Discussion}
We have shown how massive scalar hair can emerge in a dynamical situation, with non-trivial time-varying boundary conditions set by cosmology. The scalar hair grows even in simulations in which backreaction from the metric is taken into account, stabilizing into a scalar field cloud.  
The growth timescales of the cloud are comparable to the freefall timescales for the BH in high mass cases $GMm \gtrsim 1$, with the scalar cloud feeding off the reservoir at the boundary. In low mass cases $GMm \ll 1$ the growth is much more rapid, and standing wave profiles develop. Including backreaction, we find that the scalar field profile evolution is very similar, only diluted by the cosmological expansion for the values chosen -- for more realistic density values this effect would be small. We find a profile of oscillating deviations to the Ricci scalar with a scale set by $m$, which propagate outwards from the BH over time. 

To illustrate the potential cloud sizes, consider the example case of a solar mass BH in a background of DM with density $\sim M_\odot / {\rm pc}^3$ and a homogeneously oscillating region of size $L \sim {\rm pc}$. We assume that, as in our simulations, the region outside of this area is oscillating homogeneously, providing an (in principle) infinite source of scalar particles which feeds the BH and its cloud. We have seen that the energy density redistributes into a $r^{-3/2}$ profile in a time related to the freefall time $t_{\text{ff}} \sim\left( \frac{L}{R_s}\right)^{3/2} R_s$ (or faster), with the field profile matching the amplitude imposed at the boundary. We can thus quantify the time to reach a final state as $\sim10$ million years, with $\rho$ at the horizon ultimately enhanced as:
\begin{equation}
\rho = \rho_0 \left( \frac{L}{R_s} \right)^{3/2},
\end{equation}
that is, by a factor of $10^{19}$ in our example case. The equivalent figures for a supermassive BH like the one in M87, with $M \sim 10^{10} M_\odot$ in a kpc sized region, is an enhancement of $10^{9}$ on a timescale of $\sim 1$ million years. Here we neglect the impact of the growing mass of the BH due to accretion of the scalar during this period, which would not materially change the magnitude of the final state, although note that the BH can grow by a significant fraction of its initial mass in this time, as calculated in \cite{Lam2019}. The size of the induced curvature from this energy density is still, however, extremely small in comparison to the curvature of the BH, with $\rho \sim {\rm kg ~ m^3}$ even in the more favourable solar mass case.

Whilst the scalar DM case is one of the best motivated applications of these studies, we note that our results would also apply to non-DM cosmological scalars, for example, a scalar that is part of a modified gravity model. Even very tiny coherent oscillations, on cosmological scales, would build up in amplitude around BHs in a similar way to what we describe here, leading to modifications to GR localised around BHs which increase over time. We intend to study such effects further for specific models in future work.

We emphasise that our discussion neglects angular momentum, which should reduce the infall and change the picture for a more physically realistic scenario. We would still expect similar clouds to build up, only with an angular spatial dependence in the profile related to the relevant spherical harmonics. The angular momentum may also permit the clouds to grow larger, since in principle it would provide additional support against infall.

While we are interested in the end state of the process -- a quasi-stable configuration of scalar hair -- our primary interest has been in how the hair is formed dynamically and whether backreaction affects can spoil this build up in high density regions. With new windows on the Universe -- the detection of gravitational waves by Advanced LIGO/VIRGO \cite{LIGO, Virgo}, and the imaging of the event horizon of the BH in M87 \cite{2019ApJ...875L...1E} -- we are probing BHs in ever more dynamical situations. In this context, the impact of non trivial scalar clouds on a merger (see e.g.\ \cite{Horbatsch:2011ye, Wong:2019yoc, Berti:2013gfa}) merit further investigation, particularly where the binary systems have timescales similar to $1/m$ where one expects interesting resonances (see e.g.\ \cite{Blas:2016ddr,Ferreira:2017pth,Rozner:2019gba}). With the tools developed in this work, we plan to explore these questions in the future.

\section*{Acknowledgments}
\vspace{-0.2in}
\noindent We thank Vitor Cardoso for his comments on an initial draft, and the authors of \cite{Lam2019} for sharing details of their work. This project has received funding from the European Research Council (ERC) under the European Union’s Horizon 2020 research and innovation programme (grant agreement No 693024). PGF acknowledges support from STFC, the Beecroft Trust and the ERC. ML was supported by the Kavli Institute for Cosmological Physics at the University of Chicago through an endowment from the Kavli Foundation and its founder Fred Kavli.
KC is supported by the ERC. She thanks her GRChombo collaborators and acknowledges useful conversations with J Aurrekoetxea, T Helfer, EA Lim, M Radia and H Witek.

The simulations presented in this paper used the Glamdring cluster, Astrophysics, Oxford, and DiRAC resources under the projects ACLP151 and ACSP191. This work was performed using the Cambridge Service for Data Driven Discovery (CSD3), part of which is operated by the University of Cambridge Research Computing on behalf of the STFC DiRAC HPC Facility (www.dirac.ac.uk). The DiRAC component of CSD3 was funded by BEIS capital funding via STFC capital grants ST/P002307/1 and ST/R002452/1 and STFC operations grant ST/R00689X/1. The work also used the DiRAC@Durham facility managed by the Institute for Computational Cosmology on behalf of the STFC DiRAC HPC Facility (www.dirac.ac.uk). The equipment was funded by BEIS capital funding via STFC capital grants ST/P002293/1 and ST/R002371/1, Durham University and STFC operations grant ST/R000832/1. DiRAC is part of the National e-Infrastructure.

\bibliography{RefModifiedGravity}


\end{document}